\documentclass[11pt,aps,prd,preprint,preprintnumbers,showpacs,superscriptaddress,floatfix]{revtex4-1}
\usepackage{graphicx}
\usepackage{epsfig}
\usepackage{dcolumn}
\usepackage{bm}
\usepackage{subfigure}
\usepackage{amsmath}
\usepackage{amssymb}
\usepackage{slashed}
\usepackage[T1]{fontenc}
\usepackage[latin1]{inputenc}
\usepackage{caption}
\usepackage{bbm}
\usepackage{multirow}
\usepackage{setspace}
\usepackage{xcolor}
\usepackage[colorlinks,
            linkbordercolor={0 0 0},
            pdfborder={0 0 0},
            linkcolor=blue,
            citecolor=blue,
            urlcolor=blue,
            breaklinks=true]{hyperref}

\allowdisplaybreaks

\newlength{\x}
\newlength{\y}
\newlength{\z}
\urldef\milcurl\url{http://www.physics.utah.edu/~detar/milc/}

\phantomsection

\begin{document}
\preprint{IMSc/2021/02/02}

\title{$Z_N$ symmetry in $SU(N)$ gauge theories}

\author{Minati Biswal}
\email{biswalmnt@gmail.com}              
\affiliation{Indian Institute of Science Education and Research, Mohali
140306, India}

\author{Sanatan Digal}
\email{digal@imsc.res.in}              
\affiliation{The Institute of Mathematical Sciences, Chennai 600113, India}
\affiliation{Homi Bhabha National Institute, Training School Complex,
Anushakti Nagar, Mumbai 400085, India}

\author{Vinod Mamale}
\email{mvinod@imsc.res.in}
\affiliation{The Institute of Mathematical Sciences, Chennai 600113, India}
\affiliation{Homi Bhabha National Institute, Training School Complex,
Anushakti Nagar, Mumbai 400085, India}

\author{Sabiar Shaikh}
\email{sabiarshaikh@imsc.res.in}
\affiliation{The Institute of Mathematical Sciences, Chennai 600113, India}
\affiliation{Homi Bhabha National Institute, Training School Complex,
Anushakti Nagar, Mumbai 400085, India}

\begin{abstract}
  
We study $Z_N$ symmetry in $SU(N)$ gauge theories in the presence of matter fields in the 
fundamental representation, 
by restricting the lattice partition function integration to matter fields which are uniform in 
spatial directions and gauge fields with vanishing spatial components. In this approximation the
gauge matter field interaction effectively reduces  to a 1-dimensional gauged chain. This makes
analytical calculations of the matter field contribution to the Polyakov loop free energy possible. We show 
that in the limit of large number of temporal sites the explicit breaking of $Z_N$ symmetry in this
free energy vanishes, 
driven by dominance of the density of states. We argue that the spatial links as well as the spatial 
modes of the matter fields determine the boundaries separating regions where $Z_N$ symmetry 
is realised from rest of the phase diagram.
\end{abstract}


\maketitle

\section{Introduction} 
\label{sec:intro}  
Gauge theories such as quantum chromodynamics (QCD), the standard model (SM) etc. are crucial for
understanding evolutions of the early universe as well as the quark-gluon plasma (QGP) formed in 
relativistic heavy-ion collisions. Studies of phases and phase transitions in these theories are 
important as the main aim of experimental observations is to look for their signatures. One of the phase 
transitions common to all these theories is the confinement deconfinement transition(s) (CD) at 
finite temperatures. In the pure gauge limit of these theories the CD transition is described by
the center $Z_N\in SU(N)$ symmetry and the Polyakov loop which plays the role of an order 
parameter~\cite{Kuti:1980gh, Creutz:1980zw, McLerran:1980pk, Weiss:1980rj, Polonyi:1982wz, Svetitsky:1982gs, 
Yaffe:1982qf, Green:1983sd, Celik:1983wz}. Above the critical temperature, the Polyakov loop 
acquires a non-zero thermal average value which leads to the spontaneous symmetry breaking (SSB) of 
the $Z_N$ symmetry~\cite{Weiss:1980rj, Yaffe:1982qf, Celik:1983wz, Svetitsky:1985ye}. The SSB leads to $N$ 
degenerate states in the deconfined phase and global topological defects such as strings 
and domain walls in physical space~\cite{Gross:1980br, Weiss:1981ev, Balachandran:2001qn, Gupta:2010pp, 
Ignatius:1991nk}.\\

\par The $Z_N$ symmetry arises from the fact that the allowed gauge transformations in Euclidean 
space are periodic up to a factor $z\in Z_N$. These gauge transformations are responsible for the
$Z_N$ symmetry in pure gauge theory. In turn the $Z_N$ symmetry plays an important role on the nature
of the CD transition~\cite{Kogut:1979wt}. The presence of matter fields in the above gauge theories spoil 
the $Z_N$ symmetry. The requirement that the matter fields satisfy either periodic or anti-periodic boundary conditions 
in the temporal direction forces the gauge transformations to be periodic~\cite{Weiss:1981ev, 
Belyaev:1991np, Green:1983sd, Karsch:2000zv, Biswal:2015rul, Biswal:2016xyq}. However gauge field configurations related by 
$Z_N$ gauge transformations can contribute to the partition function, though not equally. Since the matter fields 
cannot be subjected to these gauge transformations the different configurations will have different actions. 
This situation appears similar to Ising model in the presence of external field~\cite{Banks:1983me}. In the 
present context, the gauge-matter field interaction plays the role of explicit $Z_N$ symmetry breaking 
term~\cite{Banks:1983me, Green:1983sd}.\\

\par There are several studies of explicit breaking of $Z_N$ symmetry due to matter fields over the 
years~\cite{Weiss:1981ev, Green:1983sd, Belyaev:1991np, Ignatius:1991nk, Dixit:1991et, Deka:2010bc, 
Biswal:2019xju}. Perturbative one loop calculations show that the $Z_N$ symmetry explicit breaking increases 
with decrease in mass of the matter fields~\cite{Weiss:1981ev}. The explicit breaking is found to increase 
with temperature. Recently there are extensions of loop calculations to higher 
order with similar trend~\cite{Guo:2018scp}. The non-perturbative studies, which are mostly around the CD 
transition regions, show decrease in explicit breaking with the number of temporal lattice points ($N_\tau$)
~\cite{Satz:1985js, Biswal:2015rul, Biswal:2016xyq}. These studies show a trend that in the continuum limit likely there 
will be reemergence of $Z_N$ symmetry.\\
The exact calculation of the partition function to validate the reemergence of $Z_N$ symmetry
is almost impossible. An explanation of this analytically, is highly desirable even with some simplifications. 
In this paper, we attempt an exact calculation of the lattice partition function after restricting the spatial
gauge fields to zero and matter fields uniform in the spatial directions. With these restrictions only the terms 
of the action which explicitly break the $Z_N$ gauge transformations remain and the problem effectively reduces
to a 1-dimensional model. The matter fields can then be integrated out exactly after a suitable gauge choice 
for the temporal links. The integrations can be carried out for any arbitrary value of $N_\tau$. From these 
calculations the free energy for the given gauge field background or Polyakov loop is obtained. The results 
show that the explicit symmetry breaking vanishes in the limit of large $N_\tau$ even when the relevant 
couplings are finite and the action breaks the $Z_N$ symmetry. These results suggest that for the parameters 
for which the $Z_N$ symmetry realised, the free energy is dominated by the density of states. Recent 
studies of $Z_2+$Higgs have shown that the histogram of the explicit breaking term exhibits $Z_2$ symmetry~\cite{Biswal:2021fde}. 
The modes/aspects of the fields neglected here will affect the $Z_N$ symmetry 
in parts of the phase diagram by driving the  system away from the point where the density of states dominate the 
thermodynamics. This is observed in Monte Carlo simulations of the partition function with the full action
~\cite{Biswal:2021fde}.\\
The paper is organised as follows. In section II, we discuss the $Z_N$ symmetry in the presence of matter fields 
in the fundamental representation. In section III, we calculate the partition function for $SU(N)+$Higgs which is
followed by the calculations for $SU(N)+$fermions in section IV. 
The discussions and conclusions are presented 
in section V.

\section{$Z_N$ symmetry in $SU(N)$ gauge theories}
The action for a minimally coupled $SU(N)$ gauge theory of fermions and bosons in $3+1$ Euclidean space is given by
\begin{eqnarray}
&&S = \int_V d^3x\int_0^\beta d\tau \left[  {1\over 2}\left\{
Tr\left(F^{\mu\nu}F_{\mu\nu}\right) + |D_\mu\Phi|^2+{m_b^2} \Phi^\dag\Phi\right\} + {\bar{\Psi}} (\slashed{D}+m_f)\Psi
\right]
\label{action1}
\\
&&F_{\mu\nu}=\partial_\mu A_\nu-\partial_\nu A_\mu + g[A_\mu,A_\nu],~D_\mu\Phi=(\partial_\mu+ ig A_\mu)\Phi,~
\slashed{D}\Psi = (\slashed{\partial} + ig\slashed{A})\Psi.\nonumber
\end{eqnarray}
Here $A_\mu$, $\Phi$ and $\Psi$ are the gauge, Higgs and the fermion fields respectively. $\Phi$ and $\Psi$ are in the fundamental
representation. $g$ is the gauge coupling strength, $m_b(m_f)$ is the mass of $\Phi(\Psi)$ fields and $\beta$ is the inverse of temperature, i.e $\beta=1/T$. The corresponding partition function takes the form
\begin{equation}
	{\cal Z} = \int [DA] [D\Phi][D\Phi^\dagger] [D\Psi] [D\bar{\Psi}]~\rm{Exp}[-S].
\end{equation}
The fields contributing to the partition function satisfy the following temporal boundary conditions,
\begin{equation}
A_\mu(\tau=0)=A_\mu(\tau=\beta),~\Phi(\tau=0)=\Phi(\tau=\beta),\Psi(\tau=0)=-\Psi(\tau=\beta).
\label{bc}
\end{equation}
These fields transform under a gauge transformation $V\in SU(N)$ as 
\begin{equation}
A_\mu\longrightarrow V A_\mu V^{-1} + {1\over g} (\partial_\mu V)V^{-1},~\Phi \longrightarrow V\Phi,~\Psi \longrightarrow V\Psi.
\end{equation}
In absence of the matter fields one can consider $V$ which is not necessarily periodic, i.e $V(\tau=0)=zV(\tau=\beta)$, where 
$z$ is an element of the center $Z_N$ of $SU(N)$. While the pure gauge action is invariant under this transformation, 
the Polyakov loop,
\begin{equation} 
	L(\vec{ x})={1 \over N} {\rm{Tr}}\left[{\rm{P}}\left\{{\rm Exp}{\left(-ig\int_0^\beta  
A_0 d\tau\right)}\right\}\right], 
\end{equation} 
transforms as $L\longrightarrow zL$. This transformation property of the Polyakov loop is crucial for it playing the role of an
order parameter for the CD phase transition and also SSB of the $Z_N$ symmetry in the deconfined phase. 
In the presence of the matter fields ($\Phi,\Psi$), the boundary conditions in Eq.\ref{bc}, restrict the gauge transformations to
be periodic in $\tau$. Since the non-periodic gauge transformations can not act on the matter fields, two configurations
with Polyakov loops $L(\vec{x})$ and $zL(\vec{x})$ do not necessarily contribute equally to the partition function. Hence
the $Z_N$ symmetry is explicitly broken. If $L(\vec{x})$ belongs to the identity sector of $Z_N$ then the configuration 
corresponding to $zL(\vec{x})$ will have higher action. 

Symmetry in the action automatically leads to symmetry in the free energy. However explicit breaking at the level of action 
does not necessarily mean the same is true at the level of free energy. This is because the free energy difference
between $L(\vec{x})$ and $zL(\vec{x})$ can only be decided after the matter fields are integrated out. Integrating the
matter field is very difficult task in four dimensions. Therefore in the following we consider $SU(N)$ gauge theory of Higgs and 
fermions separately by restricting to the spatial gauge fields and spatial variations of the matter fields to zero. As mentioned
previously in this
approximation the part of the action which breaks the $Z_N$ symmetry remains, while the system effectively reduces to
$1-$dimensional model of gauged chain, making analytical calculations possible.

\section{Gauged $1-d$ chain of $SU(N)+$Higgs}

The $SU(N)+$Higgs part of Eq.\ref{action1} on the Euclidean lattice can be written as \cite{Biswal:2016xyq},
\begin{equation}
S= \beta_g\sum_p \left[1-{1 \over 2}(U_p +U^{\dagger}_p)  \right] - b \sum_{n,\mu} (\Phi_{n}^{\dagger}U_{n,\mu}\Phi_{n+\hat{\mu}}+h.c)
+a\sum_n \Phi^\dagger_n\Phi_n.
\end{equation}
$\beta_g$ is the gauge coupling constant, $a={1 \over 2}$ and the coupling $b=(m_b^2 + 8)^{-1}$, the Higgs mass $m_b$ is expressed in lattice units~\cite{Biswal:2016xyq}.
For unit spatial links and $\Phi$ uniform in the spatial directions, the action reduces to,
\begin{equation}
S=a\sum_{i=1}^{N_\tau}\Phi_{i}^{\dagger}\Phi_{i}-b\sum_{i=1}^{N_\tau}(\Phi_{i}^{\dagger}U_{i}\Phi_{i+1} + h.c.),
\label{1dh}
\end{equation}
apart from an overall spatial volume factor. The pure gauge part is dropped as the effect of matter fields on the $Z_N$ symmetry
in the back ground of temporal gauge links is being considered.  For convenience the subscripts of the field variables have been replaced by $i$. $N_\tau$ denotes the number of temporal sites.
 $\Phi$ satisfies periodic boundary condition, i.e $\Phi_{N_\tau+1}=\Phi_1$. We consider a gauge choice in which 
$U_i=\mathbb{I}$ for $i=1,2,...,N_\tau-1$ and $U_{N_\tau}=U$. The Polyakov loop is $L=Tr(U)/N$. 
In order to derive the free energy $V(L)$, only the $\Phi_i$ fields in the partition function ${\cal{Z}}_L$ are to 
be integrated out,
\begin{equation}
	{\cal{Z}}_L = \int \prod_{i=1}^{N_\tau} d\Phi_i^\dagger d\Phi_i~{\rm Exp}[-S].
\end{equation}
For convenience the action is written as $S=S_1+S_2$ as in the following,
\begin{equation}
S_1 =  a\Phi_{1}^{\dagger}\Phi_{1}  - b\left(\Phi_{N_\tau}^{\dagger}U\Phi_{1} + h.c.\right),~
S_2=a\sum_{i=2}^{N_\tau}\Phi_{i}^{\dagger}\Phi_{i}-b\sum_{i=1}^{N_\tau-1}(\Phi_{i}^{\dagger}\Phi_{i+1} + h.c.).
\end{equation}
At first, the fields $\Phi_2$ to $\Phi_{N_\tau-1}$ are integrated out sequentially, i.e,
 \begin{equation}
	 {\cal{Z}} = \int \prod_{i=2}^{N_\tau-1} d\Phi_i^\dagger d\Phi_i~{\rm Exp}[-S_2]
\label{I9}
\end{equation}
Afterwards the remaining integration of $\Phi_1$ and $\Phi_{N_\tau}$ can be carried out to obtain the
partition function,
\begin{equation}
{\cal{Z}}_L = \int d\Phi_1^\dagger d\Phi_1d\Phi^\dagger_{N_\tau} d\Phi_{N_\tau} \left({\cal{Z}} \times {\rm Exp}[-S_1]\right).
\end{equation}
The integration of $\Phi_1$ and $\Phi_{N_\tau}$ requires evaluation of determinant of a matrix of size $4 N\times 4 N$.
The integrations of $\Phi_2$ to $\Phi_{N_\tau-1}$ greatly simplify the problem, otherwise one would have to deal with
evaluating of matrix whose size depends on $N_\tau$.
\par In the integration of ${\cal{Z}}$ in Eq.\ref{I9}, due to the gauge choice mentioned above the different components as 
well as the real and imaginary parts of $\Phi_i$'s do not mix. Therefore it can be written as,
\begin{equation}
{\cal{Z}}= \prod_{r=1}^{2N}{\cal{I}}(\Phi_{N_\tau,r}),
\end{equation}
where $\Phi_{_{N_\tau,r}}$ is the r-th component of $\Phi_{N_\tau}$ and 
 ${\cal{I}}(\Phi_{N_\tau,r})$ is obtained by integrating out r-th component of $\Phi_2$ to $\Phi_{N_\tau-1}$. Denoting the
 r-th component by $\phi$ we can write
\begin{equation}
	{\cal{I}}(\phi_{N_\tau})=\int \prod_{i=2}^{N_\tau-1} d\phi_i~{\rm Exp}[-S_2^\prime],
\label{I2}
\end{equation}
where
\begin{equation}
S_2^\prime=a\sum_{i=2}^{N_\tau -1}\phi_{i}^2-2b\sum_{i=1}^{N_\tau -1}\phi_{i}\phi_{i+1} .
\end{equation}
The integration ${\cal{I}}$, in Eq.\ref{I2}, can be also be written as,
\begin{equation}
{\cal{I}}(\phi_{N_\tau})=\int \prod_{i=3}^{N_\tau-1}d\phi_i e^{-S_3^\prime}\int d\phi_2 ~{\rm Exp}\left[-a\phi_2^2 + 2\phi_2\left(b\phi_1+b\phi_3\right)\right]
\label{I3}
\end{equation}
$S_3^\prime$ is obtained by taking out terms from $S_2^\prime$ which are dependent on $\phi_2$. After $\phi_2$ is integrated
out,
\begin{equation}
{\cal{I}}(\phi_{N_\tau})=\int \prod_{i=3}^{N_\tau-1}d\phi_i e^{-S^\prime_3}\sqrt{\pi \over a}~{\rm Exp}\left[{1\over a}\left(b\phi_1+b\phi_3\right)^2\right]
\end{equation}
which can also be written as
\begin{eqnarray}
{\cal{I}}(\phi_{N_\tau})&=&\int \prod_{i=4}^{N_\tau-1}d\phi_i e^{-S^\prime_4}\nonumber\\
&\times&\sqrt{\pi \over a} \int d\phi_3~{\rm Exp}\left[-\left(a-{b^2\over a}\right)\phi_3^2 + 2\phi_3\left({b^2\over a}\phi_1 + b\phi_4\right) + {b^2\over a}\phi_1^2\right].
\label{I4}
\end{eqnarray}
Here again $S_4^\prime$ is $S_2^\prime$ without terms containing $\phi_2$ and $\phi_3$. Given the forms of ${\cal{I}}(\phi_{N_\tau})$ in Eq.\ref{I3}
and Eq.\ref{I4} one easily write down the would be form of ${\cal{I}}(\phi_{N_\tau})$ after integration of $\phi_{k-1}$ as,
\begin{eqnarray}
{\cal{I}}(\phi_{N_\tau})&=&\int d\phi_{k+1} ....d\phi_{N_\tau -1} e^{-S^\prime_{k+1}} \nonumber\\
     &\times& I_k\int d\phi_{k}~{\rm Exp}\left[-A_{k}\phi_{k}^2 + 2\phi_{k}\left(B_{k}\phi_1 + b\phi_{k+1}\right) + E_{k}\phi_1^2\right].
\label{I5}
\end{eqnarray}
Carrying out the $\phi_{k}$ integration will result in,
\begin{eqnarray}
{\cal{I}}(\phi_{N_\tau})&=&\int d\phi_{k+2} ....d\phi_{N_\tau -1} e^{-S^\prime_{k+2}}\nonumber\\
     &\times& I_{k+1}\int d\phi_{k+1}~{\rm Exp}\left[-A_{k+1}\phi_{k+1}^2 + 2\phi_{k+1}\left(B_{k+1}\phi_1 + b\phi_{k+2}\right) + E_{k+1}\phi_1^2\right].
\label{I6}
\end{eqnarray}
From equations \ref{I5} and \ref{I6}, one can read off the following recursion relations,
\begin{equation}
I_{k+1}=\sqrt{\pi \over A_k},~A_{k+1}=a-{b^2\over A_k},~B_{k+1}={bB_k\over A_k},~E_{k+1}=E_k + {B_k^2\over A_k},
\end{equation}
with $I_2=1$, $A_2=a$, $B_2=b$ and $E_2=0$.  Using these recursion relations we can workout the integration, ${\cal{I}}(\phi_{N_\tau})$ completely. Using ${\cal{I}}(\phi_{N_\tau})$'s one can write the partition function as, 
\begin{eqnarray}
{\cal{Z}}_L&=&Q\int d\Phi_{1}^\dagger d\Phi_1 d\Phi_{N_\tau}^\dagger d\Phi_{N_\tau} \nonumber\\
&&~\rm Exp\left[-A_{N_\tau}\Phi_{N_\tau}^{\dagger}\Phi_{N_\tau}-C_{N_\tau}\Phi_{1}^{\dagger}\Phi_{1}+\left(\Phi_{N_\tau}^{\dagger}(B_{N_\tau}\mathbb{I} + bU)\Phi_{1} + H.C.\right)\right],
\end{eqnarray}
where
\begin{equation}
Q = \prod_{k=2}^{N_\tau} I_k^{^n},~ n=2N.
\end{equation}
$n$ corresponds to the number of  components of the $\Phi$ field. The coefficient $C_{N_\tau} = a - E_{N_\tau}$. After the integration of the remaining fields $\Phi_1$ and $\Phi_{N_\tau}$ the partition function takes the form,
\begin{equation}
{\cal{Z}}_L = Q \sqrt{\pi^8 \over Det(M)}.
\end{equation}
$M$ is $(4N \times 4N)$  given by,
\begin{center}
$\begin{pmatrix}
A_{N_\tau} & B_{N_\tau}+bU\\
B_{N_\tau}+bU^\dagger & C_{N_\tau}
\end{pmatrix}$	
\end{center}
The exact form of $Det(M)$ for arbitrary $N$ is difficult to find. However the sequential integration has greatly simplified
the problem. For arbitrary $N_\tau$ we need to deal with a matrix of finite size. In the following we consider $N=2$ and evaluate ${\cal{Z}}_L$ explicitly for an arbitrary $U$, which can be
parametrised as,
\begin{equation}
U = \alpha_0 + i {\bf \alpha.\sigma},~\alpha = (\alpha_1,\alpha_2,\alpha_3),
\end{equation}
where $\sigma_i$'s are the Pauli matrices. The corresponding matrix $M$ is given by,

\begin{center}
$\begin{pmatrix}
A_{N_\tau} & 0 & 0 & 0 & B_{1} & b\alpha_{3} & -b\alpha_{2} & b\alpha_{1}\\
0 & A_{N_\tau} & 0 & 0 & -b\alpha_{3} & B_{1} & -b\alpha_{1} & -b\alpha_{2}\\
0 & 0 & A_{N_\tau} & 0 & b\alpha_{2} & b\alpha_{1} & B_{1} & -b\alpha_{3}\\
0 & 0 & 0 & A_{N_\tau} & -b\alpha_{1} & b\alpha_{2} & b\alpha_{3} & B_{1}\\
B_{1} & -b\alpha_{3} & b\alpha_{2} & -b\alpha_{1} & C_{N_\tau} & 0 & 0 & 0\\
b\alpha_{3} & B_{1} & b\alpha_{1} & b\alpha_{2} & 0 & C_{N_\tau} & 0 & 0\\
-b\alpha_{2} & -b\alpha_{1} & B_{1} & b\alpha_{3} & 0 & 0 & C_{N_\tau} & 0\\
b\alpha_{1} & -b\alpha_{2} & -b\alpha_{3} & B_{1} & 0 & 0 & 0 & C_{N_\tau}
\end{pmatrix}$,
\end{center}
where $B_1= -(b\alpha_0 + B_{N_\tau})$. The  determinant of $M$ is given by,
\begin{equation}
Det{M} = \left(B^2_{N_\tau} - A_{N_\tau} C_{N_\tau} + 2 b B_{N_\tau} \alpha_0 + b^2\right)^4
\label{eq1}
\end{equation}
$Z_2$ rotation of the Polyakov loop changes $\alpha_0 \to -\alpha_0$. So in the determinant the explicit symmetry 
breaking of $Z_2$ is $2bB_{N_\tau}\alpha_0$. It is observed that $B_{N_\tau}$ rapidly decreases, vanishing in the larger $N_\tau$ 
limit restoring the $Z_2$ symmetry. Even for higher $N$ one can see the realisation of $Z_N$ symmetry as the
off diagonal elements of the matrix $M$ turn out to be just $U$ and $U^\dagger$ due to vanishing of $B_{N_\tau}$.
Effecting a $Z_N$ transformation, ie $U\to zU$, the factor $z$ in $U$ will cancel with $z^*$ in $U^\dagger$ leaving
the determinant unchanged. \\

From the above results, it can be shown that there is realisation of $Z_N$ symmetry for fixed temperature $T$ and the Higgs mass. The number of temporal sites $N_\tau$ and $T$ are related by $T = 1/(a_lN_\tau)$, where $a_l$ is the lattice
spacing . Fixing $T$ amounts to  $a_l \propto 1/N_\tau$. $a_l$ enters into the calculations
through the parameter $b=(m_b^2 + 8)^{-1}$ which depends on the Higgs mass, $m_b$, expressed in lattice units. For a fixed Higgs physical
mass, the corresponding mass in lattice units must decrease, i.e $m_b \propto a_l$. Consequently, in the $N_\tau \to \infty$ limit the dimensionless parameter $b$ increases to $b=1/8$. This increase, however, does not affect the realisation of $Z_N$ symmetry, since $bB_{N_\tau}$(Eq.\ref{eq1}) vanishes in the $N_\tau \to \infty$ limit.\\

In the following we consider the effects of staggered fermion fields on the $Z_N$ symmetry.
\vskip-2.0in
\section{ Gauged $1-d$ chain of $SU(N)+$fermions}

The lattice action for $SU(N)$ staggered fermions is given by~\cite{Kilcup:1986dg}
\begin{equation}
	S = \beta_g\sum_p \left[1-{1 \over 2}(U_p +U^{\dagger}_p)  \right]+ 2m_f\sum_{n} \bar{\Psi}_n\Psi_n + \sum_{n,\mu} \eta_{n,\mu} \left[\bar{\Psi}_n U_{n,\mu} \Psi_{n+\mu} - \bar{\Psi}_{n} U^\dagger_{n-\mu, \mu}\Psi_{n-\mu}\right]
\end{equation}
Here the fermion mass as well as the fields are expressed in lattice unit. The analog of Eq.\ref{1dh} in this case turns out to be,
\begin{equation}
S = 2m_f\sum_{i=1}^{N_\tau} \bar{\Psi}_i\Psi_i + \sum_{i=1}^{N_\tau-1} \left(\bar{\Psi}_i\Psi_{i+1} - \bar{\Psi}_{i+1}\Psi_i\right)
-{\bar\Psi}_{N_\tau} U\Psi_1 + \bar{\Psi}_1 U^\dagger\Psi_{N_\tau},
\end{equation}
The change in the sign of the last two terms is due to the anti-periodicity of $\Psi$.
Here we have considered the $KS$ phase $\eta_0$ to be $+1$~\cite{Susskind:1976jm, KlubergStern:1983dg}, however the results/conclusions do not depend on $\eta_0$. 
As in the previous section we work in the gauge in which all temporal links except the last one are set to identity. The last link
is denoted by $U$. The corresponding Polyakov loop is $L=Tr(U)/N$. To find out the free energy $V(L)$ we need to integrate out
only the fermion fields. For convenience we write $S = S_1 + S_2$ where
\begin{eqnarray}
S_1 = 2m_f\bar{\Psi}_1\Psi_1-{\bar\Psi}_{N_\tau} U\Psi_1 + \bar{\Psi}_1 U^\dagger\Psi_{N_\tau},\\
S_2 = 2m_f\sum_{i=2}^{N_\tau} \bar{\Psi}_i\Psi_i +  \sum_{i=1}^{N_\tau-1} \left(\bar{\Psi}_i\Psi_{i+1} - \bar{\Psi}_{i+1}\Psi_i\right).
\end{eqnarray}
Initially we integrate the fields $\Psi_2$, $\bar{\Psi}_2$ to $\Psi_{N_\tau-1}$, $\bar{\Psi}_{N_\tau-1}$ 
sequentially just as in the previous section. Afterwards $\Psi_1$, $\bar{\Psi}_1$ and $\Psi_{N_\tau}$, $\bar{\Psi}_{N_\tau}$ are integrated out to obtain the partition function,
\begin{equation}
{\cal{Z}}_L = \int d\bar\Psi_1 d\bar{\Psi}_{N_\tau} d{\Psi}_1  d\Psi_{N_\tau} ~{\rm Exp}[-S_1] {\cal{Z}},
\end{equation}
where ${\cal{Z}}$ is given by 
\begin{equation}
{\cal{Z}} = \int \prod_{i=2}^{N_\tau-1} d\bar\Psi_i d{\Psi}_i ~{\rm Exp}[-S_2].
\end{equation}
Since $S_2$ is diagonal in colour space we consider a particular colour of $\Psi_i$ and denote it by $\psi_i$. 
For this choice the relevant integral is,
\begin{equation}
{\cal{I}}_\psi = \int \prod_{i=2}^{N_\tau-1} d\bar\psi_i d{\psi}_i ~{\rm Exp}[-S_{2,\psi}].
\end{equation}
After integrating $\psi_2$, $\bar{\psi}_2$ and $\psi_3$, $\bar{\psi}_3$ the integral takes the form,
\begin{eqnarray}
{\cal{I}}_\psi && = \int \prod_{i=4}^{N_\tau-1} d\bar\psi_i d{\psi}_i  ~{\rm Exp}[-S^4_{2,\psi}]\times\nonumber  \\
 && \left[1+ 4m_f^2 - 2m_f\bar{\psi}_1\psi_1-2m_f\bar{\psi}_4\psi_4
+  \bar{\psi}_4\psi_1 - \bar{\psi}_1\psi_4 + \bar{\psi}_4\psi_4\bar{\psi}_1\psi_1\right]
\end{eqnarray}
$S^4_{2,\psi}$ is obtained by dropping terms which depend on $\psi_2$, $\bar{\psi}_2$ and $\psi_3$,
$\bar{\psi}_3$. The sequential integration $\psi_4$ up to $\psi_{N_\tau-1}$ and their conjugates leads to
\begin{eqnarray}
{\cal{I}}_\psi =  A_{N_\tau} - B_{N_\tau} \bar{\psi}_1\psi_1 - C_{N_\tau} \bar{\psi}_{N_\tau}\psi_{N_\tau} + \bar{\psi}_{N_\tau}\psi_1 + D_{N_\tau}\bar{\psi_1}
\psi_{N_\tau} + E_{N_\tau} \bar{\psi}_{N_\tau}\psi_{N_\tau}\bar{\psi}_1\psi_1
\end{eqnarray}
where the coefficients  $A_{N_\tau}$ to $E_{N_\tau}$ can be obtained by recursion as
\begin{equation}
A_{k+1} = 2m_fA_k + C_k, ~B_{k+1} = 2m_fB_k + E_k,~ C_{k+1}=A_k,~D_{k+1} = (-1)^{k},~E_{k+1}=B_k,
\end{equation}
with
\begin{equation}
A_4 = (1+4m_f^2), ~B_4 = 2m_f,~ C_4=2m_f,~E_4=1,
\end{equation}
Taking the ${\cal{I}}$ integrals into account we can write the partition function as
\begin{eqnarray}
{\cal{Z}}_L = \int d\bar\psi_1 d{\psi}_1 d\bar\psi_{N_\tau}  d{\psi}_{N_\tau}~\rm{Exp}\left[\bar{\psi}_{N_\tau} U \psi_1 -\bar{\psi}_1 U^\dagger\psi_{N_\tau}\right]\times\nonumber\\
\prod_r \left(1-2m_f\bar{\psi}^r_1\psi^r_1-2m_f\bar{\psi}^r_{N_\tau}\psi^r_{N_\tau} + 4m_f^2\bar{\psi}^r_1\psi_1\bar{\psi}^r_{N_\tau}\psi^r_{N_\tau}\right)\times\nonumber\\
\left(A_{N_\tau} - B_{N_\tau} \bar{\psi}^r_1\psi^r_1 - C_{N_\tau} \bar{\psi}^r_{N_\tau}\psi^r_{N_\tau} + \bar{\psi}^r_{N_\tau}\psi^r_1 + D_{N_\tau}\bar{\psi}^r_1
\psi^r_{N_\tau} + E_{N_\tau} \bar{\psi}^r_{N_\tau}\psi^r_{N_\tau}\bar{\psi}^r_1\psi^r_1\right).
\end{eqnarray}
Note that $\psi_i^r$ denotes the colour $r$ of the field $\Psi_i$ at the temporal site $i$. This expression can 
be simplified as,
\begin{eqnarray}
{\cal{Z}}_L = \int d\bar\psi_1 d{\psi}_1d\bar\psi_{N_\tau}   d{\psi}_{N_\tau}~\rm{Exp}\left[\bar{\psi}_{N_\tau} U \psi_1 -\bar{\psi}_1 U^\dagger\psi_{N_\tau}\right]\times\nonumber\\
\prod_r\left(\tilde{A} - \tilde{B}\bar{\psi}^r_1\psi^r_1 - \tilde{C} \bar{\psi}^r_{N_\tau}\psi^r_{N_\tau} + \bar{\psi}^r_{N_\tau}\psi^r_1 + \tilde{D}\bar{\psi}^r_1
\psi^r_{N_\tau} + \tilde{E} \bar{\psi}^r_{N_\tau}\psi^r_{N_\tau}\bar{\psi}^r_1\psi^r_1\right),
\label{I40}
\end{eqnarray}
where $\tilde{A}=A_{N_\tau}$, $\tilde{B}=(2m_fA_{N_\tau}+B_{N_\tau})$, $\tilde{C}=(2m_fA_{N_\tau}+C_{N_\tau})$, $\tilde{D}=D_{N_\tau}$ and $\tilde{E}=E_{N_\tau} + 2m_fC_{N_\tau} + 2m_fB_{N_\tau}+4m^2_fA_{N_\tau}$. The superscript $r$ denotes the colour of the fermion field. For large $m_f$, in ${\cal Z}_L$ the ratio of the leading and sub-leading term scales as $\sim m_f$. The sub-leading terms contain matrix elements of $U$. In the limit of $m_f\rightarrow \infty$, therefore the fermions decouple from 
the gluons and exact $Z_N$ symmetry is recovered for any $N_\tau$. For $N=2$, integration of the rest of the fields in Eq.\ref{I40} leads to,
\begin{eqnarray}
{\cal{Z}}_L &&= \tilde{E}^2 + 2\tilde{E}\tilde{A}|U_{11}|^2 + \tilde{A}^2 + 2\tilde{B}\tilde{C}|U_{12}|^2+ 2(1-\tilde{D}Re(U^2_{11}))\nonumber\\
&& +(\tilde{E}+\tilde{A})(1-\tilde{D})tr(U).
\end{eqnarray}
As one can see the $Z_2$ explicit breaking term  is linear in $\tilde{E}+\tilde{A}$. For non zero $m_f$, in the free energy $V(L)$ the first four terms of ${\cal{Z}}_L$ dominate over $\tilde{E}+\tilde{A}$. The dominance only grows with $N_\tau$, hence in the limit of large $N_\tau$ the $Z_2$ symmetry is recovered. For higher $N$ it is difficult to evaluate ${\cal{Z}}_L$ for a general $U$. To proceed further we assume the $U$ to be $U_{rs}=\lambda_r \delta_{rs}$. After the exponential in Eq.\ref{I40} is written as a polynomial,
\begin{eqnarray}
{\cal{Z}}_L &&= \int d\bar\psi_1  d{\psi}_1 d\bar\psi_{N_\tau} d{\psi}_{N_\tau}\prod_r\left(1+\lambda_r\bar{\psi}^r_{N_\tau}\psi^r_1 - \lambda^*_r\bar{\psi}^r_1\psi^r_{N_\tau} + \bar{\psi}^r_{N_\tau}\psi^r_{N_\tau}
\bar{\psi}^r_1\psi^r_1\right)\times\nonumber\\
&&\left(\tilde{A} - \tilde{B}\bar{\psi}^r_1\psi^r_1 - \tilde{C} \bar{\psi}^r_{N_\tau}\psi^r_{N_\tau} + \bar{\psi}^r_{N_\tau}\psi^r_1 + \tilde{D}\bar{\psi}^r_1
\psi^r_{N_\tau} + \tilde{E} \bar{\psi}^r_{N_\tau}\psi^r_{N_\tau}\bar{\psi}^r_1\psi^r_1\right)\\
&&=\int d\bar\psi_1 d{\psi}_1 d\bar\psi_{N_\tau}  d{\psi}_{N_\tau} \times\nonumber\\
&&\prod_r (A - B\bar{\psi}^r_1\psi^r_1 - C \bar{\psi}^r_{N_\tau}\psi^r_{N_\tau} + F_r\bar{\psi}^r_{N_\tau}\psi^r_1 + D_r\bar{\psi}^r_1
\psi^r_{N_\tau} + E_r \bar{\psi}^r_{N_\tau}\psi^r_{N_\tau}\bar{\psi}^r_1\psi^r_1),
\end{eqnarray}
where $A=\tilde{A}$, $B=\tilde{B}$, $C=\tilde{C}$, $D_r = \tilde{D}-\lambda_r^*\tilde{A}$, $E_r=\tilde{E} - \lambda_r \tilde{D}+\lambda^*_r + \tilde{A}$ and $F_r=(1+\lambda_r\tilde{A})$. After the fields are integrated out we get the following result for the partition function,
\begin{equation}
{\cal{Z}}_L = \prod_r E_r
\end{equation}
The corresponding free energy is
\begin{equation}
	V(L) = -T \sum_{r} \left\{\rm{log}\left(\tilde{E}+\tilde{A}\right) + \rm{log}\left( 1 - {\lambda_r \tilde{D}-\lambda^*_r \over \tilde{E}+\tilde{A}}\right)\right\}.
\end{equation}
For non-zero $m_f$, the second term vanishes in the limit of large $N_\tau$. Hence, in this limit, the free energy $V(L)$ is independent of $L$ leading to realisation of the $Z_N$ symmetry. The form of $U$ considered above includes $\lambda_r=\lambda$ for all $r$,
with $\lambda = ~\rm{Exp}(i2\pi n/N)$ with $n=0,1,2,...N-1$. We mention here that for this case one would have expected the explicit breaking of $Z_N$ to be maximal.

To see the realisation of $Z_N$ symmetry for a fixed temperature and physical fermion mass, the behaviour of $\tilde{E}$
and $\tilde{A}$ must be studied in the limit $N_\tau\to \infty$ while scaling the fermion mass in lattice units as $m_f\propto 1/N_\tau$. Unlike in the case of bosons, it is not possible to carry this out  analytically as the polynomial coefficients in $\tilde{E}$
and $\tilde{A}$ change with $N_\tau$. We have numerically checked that $\tilde{E}$ and $\tilde{A}$ monotonically increase with $N_\tau$ even when the lattice fermion mass scales as $m_f\propto 1/N_\tau$. The increase though is slower compared to
the case when $m_f$ is held fixed. This suggests that even for a fixed temperature and non-zero physical fermion mass there will
be realisation of $Z_N$ symmetry.

\section{Discussions and conclusions}

In this paper we report on the explicit breaking of $Z_N$ symmetry in the presence of bosons and fermions. We show that analytical treatment of the problem is possible by simplifying the functional integral where the spatial links are set to unity and matter fields uniform in spatial direction. In this simplification most of the terms of the original action drop out except for the ones which break the $Z_N$ symmetry. Also the problem reduces to 1-dimensional chain of gauged bosons/fermions making analytical calculations possible. To derive the free energy $V(L)$ for the Polyakov loop $L$, the partition function is 
evaluated for a given background of temporal gauge links. The calculations become simple in the gauge where we can set all the links except the last one to unity.  Then the matter fields are integrated out sequentially except for the two fields connected to the last gauge link. The integration of the last two fields result in determinant of a finite sized matrix for arbitrary $N_\tau$. In the case of Higgs, the $Z_N$ symmetry is realised in the partition function for $N_\tau\to\infty$. In the same limit, for fermions the explicit breaking terms drop out when the free energy $V(L)$ is calculated. The vanishingly small explicit breaking of $Z_N$
for $N_\tau\to \infty$ can be attributed to the dominance of the density of the states over the action. In this limit, the density of states is found to have the $Z_N$ symmetry ~\cite{Biswal:2021fde}. 


The present calculations leave out the effect of the spatial links and non-zero spatial modes of the matter fields. Consideration of these modes will require that spatial interaction terms be included in the action, though these terms are not directly responsible for the explicit breaking $Z_N$ symmetry.  It appears that these terms decide the onset of the explicit breaking via the Higgs and the chiral transitions, which are entropy driven. It is expected that in the phase diagram where the action dominates over the entropy or the density of states, the $Z_N$ symmetry will be explicitly broken. Recent Monte Carlo simulations in the presence of Higgs show that this is indeed the case~\cite{Biswal:2015rul,Biswal:2021fde}. The $Z_N$ symmetry is explicitly broken in the Higgs broken phase even for large $N_\tau$. 


\acknowledgements

We thank A. P. Balachandran, S. Datta and S. Sharma for valuable discussions and suggestions.

\end{document}